# Ionic Selectivity of Nanopores: Comparison among Cases under the Hydrostatic Pressure, Electric Field, and Concentration Gradient

Chao Zhang [a], Mengnan Guo [a], Hongwen Zhang [b], Xiuhua Ren [a], Yinghao Gao [a], and Yinghua Qiu [b,*]

[a] *School of Mechanical and Electronic Engineering, Shandong Jianzhu University, Jinan, 250101, China*

[b] *Key Laboratory of High Efficiency and Clean Mechanical Manufacture of Ministry of Education, National Demonstration Center for Experimental Mechanical Engineering Education, School of Mechanical Engineering, Shandong University, Jinan, 250061, China*

*Corresponding author

E-mail address: yinghua.qiu@sdu.edu.cn (Yinghua Qiu)

**ABSTRACT**

The ionic selectivity of nanopores is crucial for the energy conversion based on nanoporous membranes. It can be significantly affected by various parameters of nanopores and the applied fields driving ions through porous membranes. Here, with finite element simulations, the selective transport of ions through nanopores is systematically investigated under three common fields, i.e. the electric field (V), hydrostatic pressure (p), and concentration gradient (C). For negatively charged nanopores, through the quantitative comparison of the cation selectivity ($t_+$) under the three fields, the cation selectivity of nanopores follows the order of $t_{+V} > t_{+c} > t_{+p}$. This is due to the transport characteristics of cations and anions through the nanopores. Because of the strong transport of counterions in electric double layers under electric fields and concentration gradients, the nanopore exhibits a relatively higher selectivity to counterions. We also explored the modulation of $t_+$ on the properties of nanopores and solutions. Under all three fields, $t_+$ is directly proportional to the pore length and surface charge density, and inversely correlated to the pore diameter and salt concentration. Under both the electric field and hydrostatic pressure, $t_+$ has almost no dependence on the applied field strength or ion species, which can affect $t_+$ in the case of the concentration gradient. Our results provide detailed insights into the comparison and regulation of ionic selectivity of nanopores under three fields which can be useful for the design of high-performance devices for energy conversion based on nanoporous membranes.

# 1. Introduction

Mass transport inside nanoporous membranes plays a crucial role in energy conversion, including the conversion of electrical energy into mechanical energy,[1, 2] hydrostatic pressure,[3, 4] or osmotic energy into electrical energy.[5, 6] As the functional units of porous membranes, nanopores provide a versatile platform for investigating the mass transport characteristics in confined spaces.[7, 8] Under highly confined spaces such as nanopores, the transport of ions and fluid can be regulated by the properties of pore walls due to interface effects. At solid-liquid interfaces, counterions in the solution are attracted to the charged solid surface to form electric double layers (EDLs).[7] After the application of electric fields, hydrostatic pressure, or concentration gradients across the nanopore, counterions inside EDLs have a directional transport. Due to the much higher concentration of counterions inside EDLs than coions, the nanopores exhibit ionic selectivity to counterions.[9-11]

Under the electric fields, the directed migration of counterions inside EDLs can induce electroosmotic flow (EOF)[7, 11] because of the ionic hydration,[12] which converts electrical energy into kinetic energy of the fluid. The generation of EOF inside nanopores is inseparable from the ionic selectivity of nanopores. Based on EOF inside the porous membrane, the electroosmotic pump can be manufactured,[13] which has various potential applications, such as the delivery of drugs,[14] and liquids,[15] as well as nanofluidic sensors.[16] The ionic selectivity of nanopores influences the performance of the EOF pump, such as the flow rate, output flux, and output pressure.[17] Chen et al.[18] used an alumina nanoporous membrane for electroosmotic pumping at low applied voltages. They found that the maximum flow rate was 0.09 mL min$^{-1}$ V$^{-1}$ cm$^{-2}$ and the energy conversion efficiency was ~0.43%.

The applied hydrostatic pressure promotes the liquid transport through nanopores which results in the pressure-driven migration of ions.[19] With the ionic selectivity of nanopores, a potential difference can be induced across the porous membranes, i.e. streaming potentials.[20] Then, the mechanical energy is converted into electrical energy.[3, 21] The stability and efficiency of the conversion process are determined by the permeability and ionic selectivity of the nanoporous membranes.[21, 22] Under the hydrostatic pressure, van der Heyden et al.[4, 21] found that the energy conversion efficiency reached the highest value when the double electric layers overlapped in nanopores. However, with a single rectangular nanochannel of 75 nm in height, the highest energy conversion efficiency was only ~3%.[4]

Under concentration gradients, a considerable electrical potential can be induced across the porous membrane with ionic selectivity as a result of the counterion diffusion.[23, 24] This process converts the chemical energy on both sides of the membranes into electrical energy, which is known as reversed electrodialysis.[5] At estuaries, vast osmotic energy is released when fresh water in rivers mixes with seawater, which is considered a promising clean, renewable, and non-intermittent energy source. Using reversed electrodialysis the osmotic energy can be widely mined. During the process of osmotic energy conversion, both the output power and conversion efficiency depend directly on the ionic selectivity of nanopores.[25-27]

From the literature, various nanoporous membranes with high ionic selectivity have been developed to achieve high-performance osmotic energy conversion.[28-31]

In all three kinds of energy conversion based on nanoporous membranes, high ionic selectivity of nanopores is required to improve the energy conversion efficiency, which is closely related to the dimensions of nanopores, surface charge density, and properties of aqueous solutions. In previous studies, though the ionic selectivity of nanopores has been widely involved,[9, 10, 27, 32] it was usually considered under one field, such as the hydrostatic pressure, electric field, or concentration gradient. Here, COMSOL Multiphysics was used to investigate the characteristics of ionic selective transport through nanopores under all three fields. We focused on the ionic selectivity of nanopores under different fields which can be affected by the pore parameters and solution properties. Based on the detailed characteristics of ion transport across membranes, we compared the ionic selectivity of nanopores under three different fields and systematically explained the microscopic mechanism. With the increased energy demand, the research on sustainable energy and energy conversion efficiency has attracted great attention.[33] Under the rapid development of nanotechnology, micro/nanofluidics with the application of nanoporous materials have shown broad application prospects in the fields of energy conversion and collection.[5, 6] In this work, our results may guide the development of high-performance nanoporous membranes for energy conversion.

## 2. Methodology

Simulation models of ion transport in nanopores under electric fields, hydrostatic pressure, and concentration gradients were built with COMSOL Multiphysics (Fig. 1). The system consists of two reservoirs with a radius and length of 5 μm, and a cylindrical nanopore connecting both reservoirs. Coupled Poisson-Nernst-Planck (PNP) and Navier-Stocks (NS) equations were solved to consider ion distributions near charged membrane surfaces, ionic diffusion and migration in aqueous solution, as well as the movement of liquid in the system, as described by Equations 1−4.[9, 11, 34, 35]

$$\varepsilon \nabla^2 \varphi = -\rho_e = -F \sum_{i=1}^{N} z_i C_i \tag{1}$$

$$\nabla \mathbf{J}_i = \nabla \left( -D_i \nabla C_i + \mathbf{u} C_i - \frac{F z_i C_i D_i}{RT} \nabla \varphi \right) = 0 \tag{2}$$

$$\mu \nabla^2 \mathbf{u} - \nabla p - \sum_{i=1}^{N} \left( z_i F C_i \right) \nabla \varphi = 0 \tag{3}$$

$$\nabla \mathbf{u} = 0 \tag{4}$$

where $\varepsilon$, $\nabla$, $\varphi$, $\rho_e$, $F$, and $N$ are the dielectric constant of solutions, gradient operator, electrical potential, volumetric charge density, Faraday constant, and number of ionic species. $z_i$, $C_i$, $\mathbf{J}_i$, and $D_i$ are the valence, concentration, ionic flux, and diffusion coefficient of ionic species $i$ (including cations and anions). $\mathbf{u}$, $R$, and $T$ are the velocity of the fluid, gas constant, and temperature, respectively. $\mu$ and $p$ are the viscosity of solutions and pressure.

During simulations, the temperature was set to room temperature 298 K, and the relative dielectric constant of water was 80. Both reservoirs were filled with NaCl solution, and the diffusion coefficients of $Na^+$ and $Cl^-$ ions were $1.33 \times 10^{-9}$ and $2.03 \times 10^{-9}$ $m^2$ $s^{-1}$, respectively.[36] A series of simulations were conducted to investigate the dependence of ionic selectivity on the nanopore size, surface charge density, and solution properties. The pore length was considered from the nanoscale to the microscale, with a variation range of 50 nm to 5 μm. The default pore length was 1 μm. The pore diameter was varied from 4 to 20 nm with a default value of 10 nm.[11, 17, 37] The nanopore surfaces were negatively charged with a density varying from −0.005 to −0.16 C/m$^2$. −0.08 C/m$^2$ was considered as the default surface charge density.[38, 39] In the cases with pressure gradients and electric fields, the concentration of NaCl solution was changed from 0.01 to 0.5 M, and the default concentration was 0.1 M. In the cases with concentration gradients, the natural salt gradients that exist at the estuary of rivers were considered as the default concentration gradient. The NaCl solutions on the low-concentration side ($C_L$) and the high-concentration side ($C_H$) were 0.01 M and 0.5 M to mimic the river water and seawater, respectively. For other concentration gradients, $C_H$ was changed from 0.01 to 1 M, with a maximum concentration ratio of 100.[5] Due to the relatively small current obtained under pressure gradients, a hydrostatic pressure of 4 MPa was applied across the pores. For the cases with electric fields, 0.6 V was applied across the nanopores. In our simulations, fluid flow was considered which may have a significant impact on the ion transport through nanopores.[11] Table S1 lists all the boundary conditions used in the simulations.

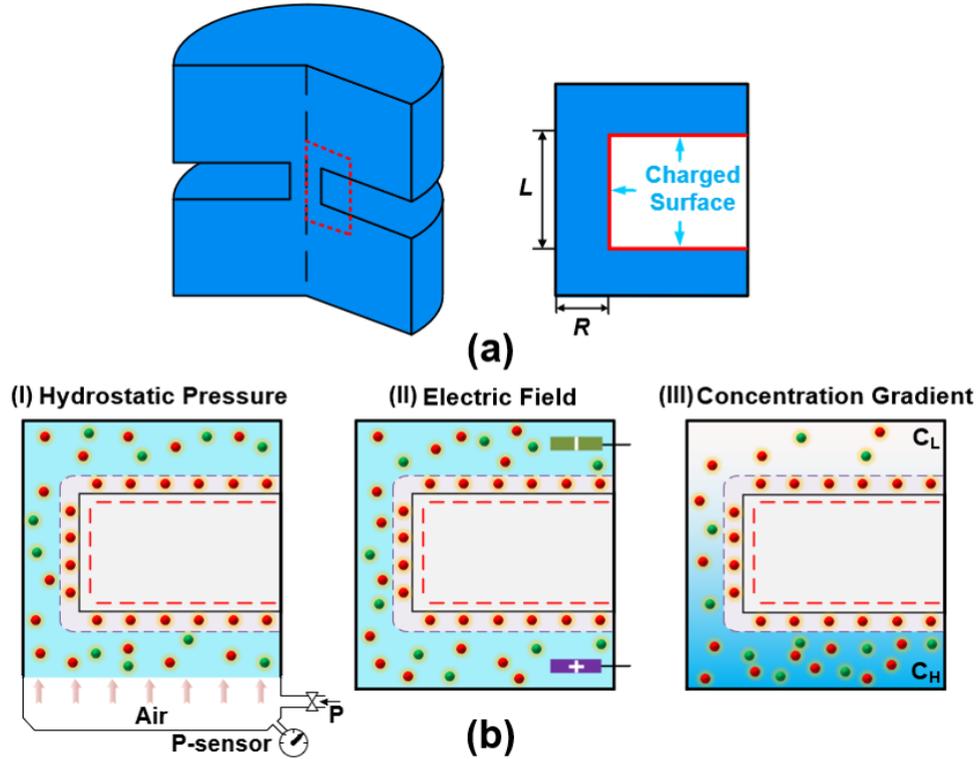

Fig. 1. Schemes of nanofluidic simulations under hydrostatic pressure, electric fields, and concentration gradients. (a) Simulation model, including two reservoirs and a nanopore. The zoomed-in part shows the nanopore region. $L$ and $R$ denote the length and radius of the nanopore. Red lines represent negatively charged pore surfaces. (b) Simulation diagram of ion and fluid transport in nanopores under three different fields, (I) hydrostatic pressure, (II) electric fields, and (III) concentration gradients. $C_H$ and $C_L$ are the concentrations of the high- and low-concentration solutions under the concentration gradient. Red and green spheres represent $Na^+$ and $Cl^-$ ions, respectively.

To explore the influence of the surface charge density on ionic selectivity, exterior surface charges were considered in our simulation models besides the inner-pore surface.[9, 34, 35] Considering the modulation of ion transport from EDLs and the thickness of the EDLs, the mesh size on the charged pore walls in our models was set to 0.1 nm, including the inner-pore surface and the exterior surfaces within 3 µm beyond the pore boundary.[9, 27, 34, 35] Due to the large reservoirs used in the system, exterior surface charges far away from the pore boundary have a weak influence on ion transport through nanopores. To reduce the calculation cost, a mesh size of 0.5 nm was adopted on the rest exterior surfaces beyond 3 µm away from the pore boundary (Fig. S1).

In our simulation cases, the ionic current ($I$) contributed by cations ($I_+$) or anions ($I_-$) through the nanopore can be obtained by integrating the ionic flux ($J_i$) at the reservoir boundary ($S$) with Equation 5.[9, 34, 40]

$$I_{\pm} = \int_S F z_i \boldsymbol{J}_i \boldsymbol{n} dS \tag{5}$$

where $\boldsymbol{n}$ is the unit normal vector.

In the negatively charged nanopores, the ionic selectivity can be calculated using Equation 6, where $t_+$ is the cation transfer number.

$$t_+ = |I_+| / (|I_+| + |I_-|) \tag{6}$$

in which $|I_+|$ and $|I_-|$ are the absolute current values of cations and anions, respectively.

Note that in this work, due to the negative surface charges on pore walls, the cation selectivity ($t_+$) is analyzed,[10] which is denoted as $t_{+V}$, $t_{+p}$, and $t_{+c}$ under electric fields, hydrostatic pressure, and concentration gradients, respectively.

## 3. Results and discussion

The characteristics of ion transport inside nanopores under three different fields have been investigated, including the hydrostatic pressure, electric field, and concentration gradient. Based on the ion currents contributed respectively by anions and cations through the nanopore, we conducted a systematical study on the ionic selectivity of the nanopore,[9, 34] considering the influences of different nanopore parameters and salt solution properties.[7]

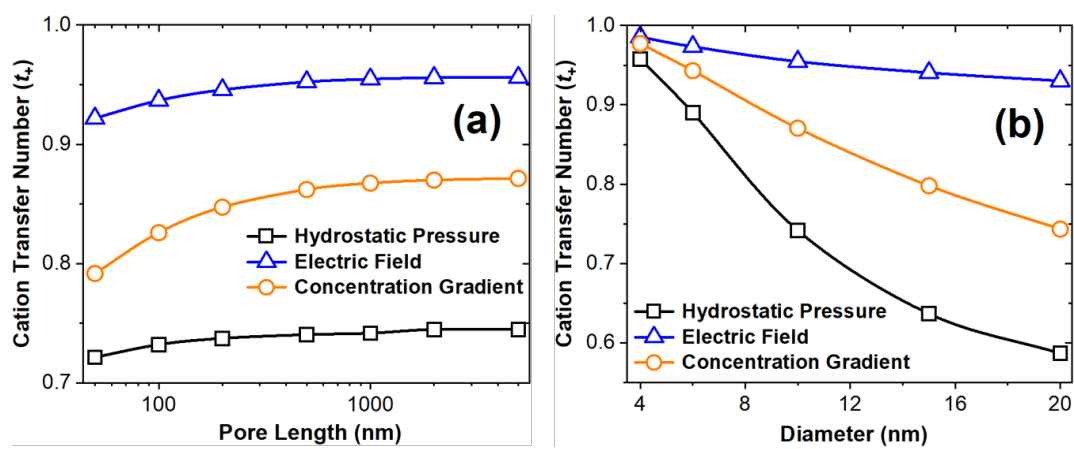

Fig. 2. Effects of the nanopore dimensions on ionic selectivity under different fields, including the pore length (a) and pore diameter (b). In the cases with different fields, default parameters were adopted. A hydrostatic pressure of 4 MPa and a voltage of 0.6 V were applied under the hydrostatic pressure and electric field, respectively. 0.1 M NaCl solution was used inside reservoirs and nanopores in the above cases. A concentration gradient of 0.01: 0.5 M NaCl solution was applied under the concentration gradient. The surface charge density of pore walls was −0.08 C/m$^2$.

Nanopores are the functional units of the porous membrane used for energy conversion. Fig. 2 shows the effects of nanopore dimensions, i.e. the pore length and diameter, on the ionic selectivity of the nanopore under the electric field, hydrostatic pressure, and concentration gradient. For 10-nm-in-diameter nanopores, from the dependence of the ionic selectivity on the pore length under different fields (Fig. 2a), the ionic selectivity of nanopores follows the order of $t_{+V} > t_{+c} > t_{+p}$ where the subscripts *V*, *c*, and *p* denote the electric field, concentration gradient, and hydrostatic pressure. With the increase of the pore length, the larger charged inner-pore surface can produce a stronger electrostatic attraction and repulsion to the mobile cations and anions, respectively, which can improve the cation selectivity of nanopores.[10, 27]

Under the three different fields, the ionic selectivity of the nanopore increases as the pore length expands until it reaches 500 nm. Under the electric field, the cation selectivity presents the highest values which stay above 0.9. In this case, the induced EOF shares the same direction as the movement of counterions, which promotes and inhibits the transport of counterions and coions through nanopores, respectively (Fig. S2a).[11] The ionic selectivity ($t_{+V}$) exhibits a slight increase from 0.92 to 0.95 as the pore length increases from 50 to 500 nm and remains at ~0.95 for nanopores longer than 500 nm. Under the concentration gradient, the ionic selectivity ($t_{+c}$) has a significant dependence on the nanopore length, which increases by 8.8% as the pore length expands from 50 to 500 nm. When the nanopore length exceeds 500 nm, the ionic selectivity of nanopores shows a weak increasing trend, which is similar to that under electric fields. Note that the induced diffusio-osmotic flow inside the nanopore decreases the cation selectivity (Fig. S2b). This is attributed to the relatively larger increase in the diffusive flux of coions by the diffusio-osmotic flow than that of cations.[27] Among the three fields, the nanopore has the weakest ionic selectivity to cations ($t_{+p}$) under hydrostatic pressure, which corresponds to the reported low efficiency during the energy conversion.[4, 21] As the pore length changes from 50 to 5000 nm, the ionic selectivity of nanopores increases from 0.72 to 0.745 by only 3.47%, indicating that it is not sensitive to the nanopore length under hydrostatic pressure.

The diameter of nanopores determines the confinement degree of the space, which affects the modulation of ion transport inside nanopores by the properties of pore walls.[7] Fig. 2b illustrates the ionic selectivity of 1000-nm-in-length nanopores with different diameters under three different fields. In the nanopore with a diameter of 4 nm, the high confinement enables the nanopore a strong selectivity to cations.[10] In this case, the ionic selectivity of nanopores shares a similar value in all three applied fields. As the nanopore diameter increases, the confinement degree of nanopores decreases which induces decreased electrostatic interaction between surface charges and ions inside nanopores. Consequently, the ionic selectivity of nanopores decreases sharply with the pore diameter expanding from 4 to 20 nm.[11] Among the three applied fields, the ionic selectivity under the hydrostatic pressure has the largest decrease from 0.958 at d=4 nm to 0.587 at d=20 nm by 38.7%. This is resulted from the significant enhancement of the flux of $Cl^-$ ions caused by the increased pressure-driven flow inside larger nanopores. Under the electric field, the ionic current is consistently dominated by $Na^+$ ions. The cation selectivity has a relatively weaker dependence on the variation of the pore diameter, which only decreases by 5.58% as the pore diameter increases from 4 to 20 nm. This may be due to the strong EOF which can promote and inhibit the transport of counterions and coions, respectively.[7, 11]

Under the concentration gradient, as the nanopore diameter increases, more counterions and anions are located away from the EDLs and have ignored electrostatic interaction with the surface charges. Then, the increase in the diffusion flux of $Cl^-$ ions results in a gradual decrease in cation selectivity which decreases from 0.98 to 0.74 by 24.5% with the pore diameter increasing from 4 to 20 nm.

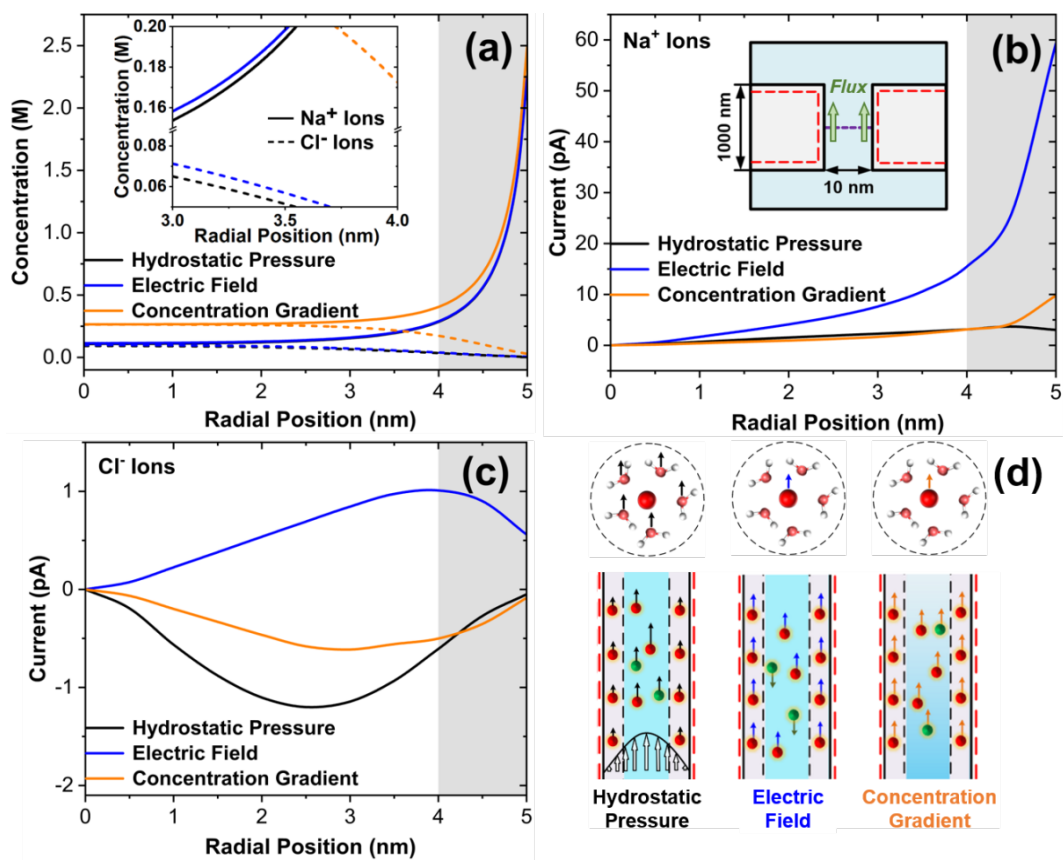

Fig. 3. Characteristics of ion transport through nanopores under different fields, including the radial distributions of ionic current and ion concentrations in the central cross-section of the nanopore. (a) Radial distributions of ion concentrations. (b-c) Ionic current contributed by $Na^+$ ions (b) and $Cl^-$ ions (c). The inset shows the location of the central cross-section. (d) Illustrations of ion transport through the nanopore under the hydrostatic pressure, electric field, and concentration gradient, respectively.

The EDLs regions near charged pore walls are shown in light grey. In the simulations, the diameter and length of nanopores were 10 nm and 1000 nm. A hydrostatic pressure of 4 MPa and a voltage of 0.6 V were applied across the nanopore in 0.1 M NaCl under a hydrostatic pressure gradient and electric field, respectively. In cases with the concentration gradient, 0.01: 0.5 M NaCl solution was applied. The surface charge density of pore walls was −0.08 C/m$^2$.

Inside charged nanopores, electrostatic interactions between surface charges and mobile ions induce the difference in the flux of anions and cations, which presents as the ionic selectivity of the nanopore.[41] Considering the axial transport direction of ions and fluid through highly confined nanopores, the radial distributions of ionic flux at the cross-section can shed light on the local contribution of ions at different radial positions to the total current.[9, 35] Also, the ion flux is determined by the radial concentration distribution of ions. So, to elucidate the characteristics of the ionic selectivity of nanopores under different fields, i.e. $t_{+V} > t_{+c} > t_{+p}$, we investigated the microscopic details of ion transport through nanopores, including the distributions of ion flux and concentration in the radial direction at the central pore cross-section.[9, 27]

Fig. 3a presents the radial concentration distributions of Na$^+$ and Cl$^-$ ions in the central cross-section. In the cases under the electric field and hydrostatic pressure, 0.1 M NaCl solution yields a Debye length of ~1 nm. Inside EDLs near the charged surface, cations and anions have aggregation and depletion, respectively, which can be described by the Poisson-Boltzmann equation.[41] Beyond ~1 nm from the pore wall, the ion concentration gradually approaches the bulk value. While under the concentration gradient, the local concentration at the central cross-section is ~0.25 M corresponding to a Debye length of ~0.6 nm.

With the application of fields across the membrane, the induced ion transport results in the ionic current through the nanopore. Figs. 3b and 3c show the radial distributions of current, i.e. the ionic flux, contributed by cations and anions at the central cross-section of the nanopore. Under the three fields, counterions in the EDLs provide a significant contribution to the cation current, especially for the cases with electric fields.[9, 35] The cation flux ($Q$) near the charged walls follows the order of $Q_{+V} > Q_{+c} > Q_{+p}$ which agrees well with that of the ionic selectivity. Because of the electrostatic repulsion between negative surface charges and anions, the transport of $Cl^-$ ions mainly happens beyond the EDLs.

Under the electric field, the applied voltage provides a significant driving force to mobile ions inside the nanopore. The transport of counterions in EDLs forms surface conductance which provides a dominant contribution to the ionic current.[9] Meanwhile, the directional migration of counterions inside EDLs can induce EOF through the nanopore, which can promote and inhibit the transport of cations and anions, respectively.[11] Due to the inhibition of EOF, the largest flux of $Cl^-$ ions appears in the region of ~2 nm to 0.5 nm away from the surface. In this case, the ionic current is dominated by counterions i.e. $Na^+$ ions. Compared with the other two cases, cations have a much larger flux under electric fields which results in the largest cation selectivity of nanopores.

Under the concentration gradient, cations share a similar current distribution profile to that in the electric field but with much lower values. In this case, the driving force for ion transport through the nanopore is the directional diffusion of ions under the concentration gradient.[27] Within 1 nm away from the inner-pore surface, EDLs provide a fast passageway for the diffusive transport of counterions.[35] Compared with the cases under electric fields, the weaker diffusive driving force results in significantly decreased cation flux values. The induced diffusio-osmotic flow by the diffusion of counterions inside EDLs can promote the diffusive transport of both cations and anions.[42] Note that under concentration gradients, both cations and anions have the same moving direction, resulting in opposite ionic currents. The strong inhibition of surface charges on anion diffusion produces greater cation selectivity, which is of great importance for improving the efficiency of osmotic energy conversion.[25, 26]

Under hydrostatic pressure, the liquid inside the nanopore forms a pressure-driven flow with a parabolic velocity profile (Fig. S3), which induces the directional movement of free ions via the ion hydration effect.[19] Both cations and anions share the same direction along the pressure gradient. Due to the regulation of the ion distributions by surface charges, the high-concentration counterions inside EDLs near the charged pore walls contribute to a large portion of the current which results in the cation selectivity of nanopores. However, in contrast to the other two cases, the motivation of ionic transport is provided by the liquid flow inside the nanopore. As shown by the parabolic velocity profile, the pressure-driven flow has the highest velocity in the nanopore center, but the lowest speed near the pore walls. At the non-slip charged surfaces, the velocity of the fluid flow is almost zero.[43] In this case, the large number of counterions inside the EDLs can not be driven effectively to produce a considerable flux near the pore walls. While in the region beyond the EDLs, $Cl^-$ ions can have a fast transport. Then, a weak selectivity to cations is induced under the hydrostatic pressure.[4, 21] Due to the weaker cation transport in EDLs and stronger anion transport in the pore center than those in the other two cases, the lowest cation selectivity of the nanopore appears under the hydrostatic pressure.

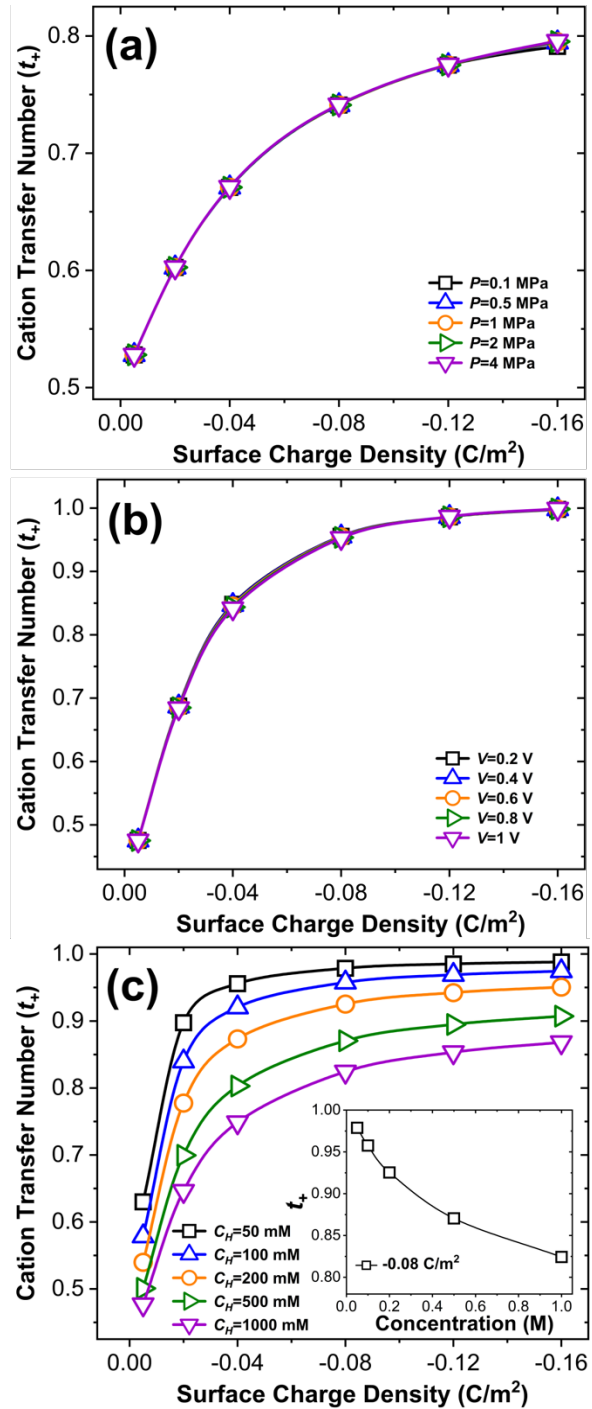

Fig. 4. Effects of the surface charge density and field strength on ionic selectivity of nanopores under the three fields. (a) Hydrostatic pressure, (b) electric field, and (c) concentration gradient. The inset in (c) shows the dependence of $t_+$ on the high concentration ($C_H$) at −0.08 C/m². The diameter and length of nanopores were 10 nm and 1000 nm. 0.1 M NaCl solution was applied under the hydrostatic pressure and electric field.

Inside charged nanopores, the surface charge density controls the strength of the electrostatic attraction exerted on mobile cations near pore walls, which determines the degree of aggregation of counterions inside the EDLs, and has a significant impact on the ion transport through the nanopore. Here, the quantitative modulation of ionic selectivity by the surface charge density is explored under the hydrostatic pressure, electric field, and concentration gradient with different field strengths. As shown in Fig. 4, under the three fields, the increased surface charge density can induce an enhancement in the cation selectivity of nanopores. This is attributed to the promoted electrostatic interaction between the pore walls and ions, which respectively enhances and inhibits the transport of cations and anions.

Figs. 4a and 4b exhibit the variation of the ionic selectivity of nanopores with the surface charge density under different hydrostatic pressures and applied voltages. It is noteworthy that in both cases, the applied strength of hydrostatic pressure and electric field has little influence on the degree of the cation selectivity of nanopores. This is mainly attributed to the independent ionic concentration distributions inside EDLs on the applied field strength (Fig. S4). In both cases, with the surface charge density increasing from −0.005 to −0.16 C/m$^2$, the cation selectivity increases from 0.53 to ~0.795 by ~50%, and from 0.475 to 0.998, by ~110%, respectively. Fig. 4c shows the dependence of cation selectivity on the surface charge density under various concentration gradients which have the same low-concentration solution of 0.01 M. The cation selectivity presents a decreasing trend with the increase of the concentration gradient. This is attributed to the better screening of surface charges by the higher salt concentration, which leads to the enhancement in the transport of $Cl^-$ ions. For cases with a surface charge density less than $-0.04$ C/m$^2$, the cation selectivity exhibits a clear sensitivity to the variation in the surface charge density.

Note that under the electric field and concentration gradient, both anions and cations share the same moving direction. In the cases with low surface charge densities, such as $-0.005$ C/m$^2$, the nanopores exhibit weak selectivity of $Cl^-$ ions due to the larger diffusion coefficient of $Cl^-$ ions than $Na^+$ ions.[27, 36]

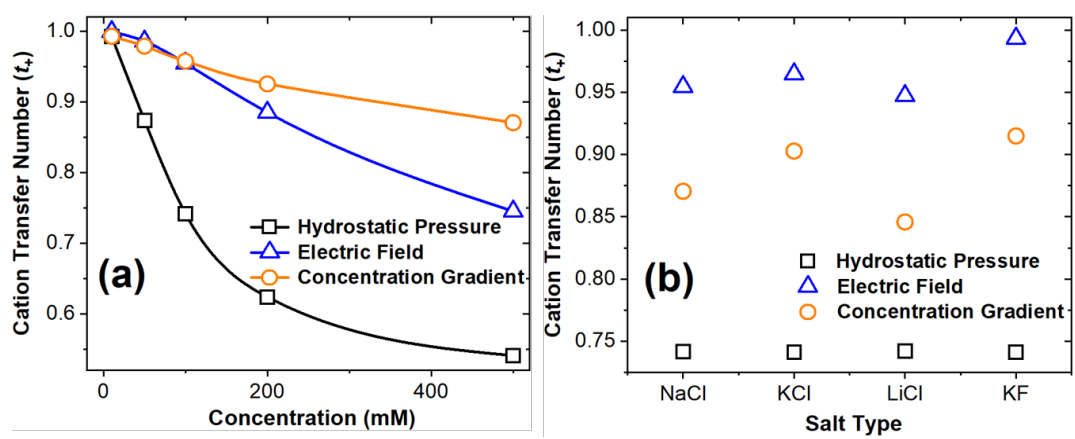

Fig. 5. Influence of solution properties on the ionic selectivity of nanopores under different fields, including salt concentration (a) and ion species (b). The pore length and diameter were 1000 and 10 nm. A hydrostatic pressure of 4 MPa and a voltage of 0.6 V were applied across the nanopore in the cases under the hydrostatic pressure and electric field, respectively. The surface charge density of pore walls was $-0.08$ C/m$^2$. In Fig. 5a, under the concentration gradient, the horizontal coordinate denotes the high concentration ($C_H$). The low concentration ($C_L$) was set at 0.01 M. In Fig. 5b, the concentration gradient was selected to the natural salt gradient at the estuary, i.e.

$C_H$ and $C_L$ were 0.5 and 0.01 M, respectively.

Considering the direct dependence of ion transport on properties of salt solution, such as the concentration and diffusion coefficient, here, we investigated the influences of salt concentration and electrolyte type on the ionic selectivity of nanopores. In aqueous solutions, the ion concentration determines the degree of counterions to shield surface charges, which can be quantitatively characterized by the Debye length.[41] In highly concentrated solutions, a small Debye length means the well screening to surface charges by counterions. As shown in Fig. 5, under the three fields, the surface charges can be better shielded by counterions in higher-concentration solutions. Due to the decreased rejection of the $Cl^-$ ion by the charged pore walls, the flux of $Cl^-$ ions increases, resulting in reduced cation selectivity of the nanopore.

From Fig. 5a, the nanopore under hydrostatic pressure presents the most significantly decreased cation selectivity at high salt concentrations. As the concentration increases from 0.01 to 0.5 M, the cation selectivity of the nanopore decreases by 45.6% from 0.992 to 0.54. Especially, in the concentration region from 0.01 to 0.2 M, the ionic selectivity exhibits a decrease of 37.1%. The nanopore becomes selective to anions in 0.5 M NaCl due to the larger diffusion coefficient of $Cl^-$ ions. Under the electric field, the ionic selectivity of nanopores has a roughly linear decrease with the salt concentration. In this case, because of the strong driving force provided by the applied electric field on mobile ions and the induced strong EOF, the decrease in the cation selectivity is smaller than that under hydrostatic pressure. With the concentration increasing from 0.01 to 0.5 M, the cation selectivity of the nanopore decreased by 25.4% from 0.999 to 0.745. Under concentration gradients, because of the 0.01 M solution on the low-concentration side, relatively thicker EDLs enable the nanopore a larger cation selectivity than the other two cases. The applied concentration gradients across the nanopore have small impacts on the ionic selectivity, which decreases from 0.98 to 0.87 as $C_H$ changes from 0.05 to 0.5 M.

The diffusion coefficient determines the transport velocity of ions through nanopores, including diffusion and migration.[44] Here, different electrolyte types such as KCl, LiCl, NaCl, and KF are applied to explore the influence of diffusion coefficients of cations and anions on the ionic selectivity of nanopores.[9] The diffusion coefficients of $K^+$, $Li^+$, and $F^-$ ions were set to $1.96 \times 10^{-9}$, $1.03 \times 10^{-9}$, and $1.475 \times 10^{-9}$ $m^2\ s^{-1}$, respectively.[36] In aqueous solutions, a larger ionic diffusion coefficient leads to higher diffusion and electro-migration rates by the Nernst-Planck equation,[44] resulting in a larger ionic flux (Fig. S5).

By analyzing the currents contributed by cations and anions under the electric field, hydrostatic pressure, and concentration gradient, Fig. 5b plots the obtained ionic selectivity of nanopores in different electrolyte solutions. Under hydrostatic pressure, due to the dominant impact of pressure-driven flow on ionic flux rather than that of ion diffusion, a weak dependence of the cation selectivity on the ion type appears. However, under the concentration gradient, the ionic flux is mainly contributed by the directional ionic diffusion through nanopores which is determined by the diffusion coefficient of cations and anions. The ionic selectivity of nanopores exhibits a strong correlation to the applied electrolyte which can be weakly affected by the diffusio-osmotic flow (Fig. S6). In the solution with a large cation diffusion coefficient and a small anion diffusion coefficient, a high cation selectivity of the nanopore can be achieved. Under the electric field, due to the strong promotion and inhibition of the EOF to the flux of cations and anions (Fig. S6), the cation selectivity stays at ~0.95 and above in the considered electrolytes. The ion type i.e. diffusion coefficient has a much weaker effect on the ionic selectivity of nanopores than that in the case under the concentration gradient.

## 4. Conclusions

With finite element simulations, the characteristics of selective ionic transport through nanopores were systematically investigated under three different fields. We also explored the dependence of the ionic selectivity of nanopores on nanopore dimensions, surface charge density of pore walls, salt concentrations, and electrolyte types. Based on our quantitative comparison of the cation selectivity among the three fields, we achieved the following main conclusions:

(1) Relative magnitude of ionic selectivity. For negatively charged nanopores, the cation selectivity ($t_+$) of nanopores follows the order of $t_{+V} > t_{+c} > t_{+p}$ where the subscripts $V$, $c$, and $p$ denote the electric field, concentration gradient, and hydrostatic pressure.

(2) Microscopic ion transport through nanopores. The ionic selectivity of nanopores is mainly caused by the microscopic transport of ions through nanopores under different fields. Under electric fields and concentration gradients, counterions inside EDLs provide a dominant contribution to the total current. However, under hydrostatic pressure, the pressure-driven flow has the weakest driving force to the counterions inside EDLs.

(3) The effect of pore parameters on ionic selectivity. Under the three fields, the ionic selectivity of nanopores is proportional to the pore length and surface charge density, and inversely proportional to the pore diameter. Compared with the pore length, the pore diameter has a more significant influence on the nanopore selectivity, especially for the cases under hydrostatic pressure.

(4) The influences of the field strength on ionic selectivity. Under the electric field and hydrostatic pressure, the ionic selectivity of nanopores has little dependence on the applied field strength, due to the unchanged ion concentration distribution inside EDLs. However, the concentration gradient has a significant effect on the ionic selectivity of nanopores. A higher concentration gradient can increase the salt concentration inside the nanopore and result in a lower ionic selectivity.

(5) The influences of solution properties on ionic selectivity. The application of high salt concentrations can decrease the ionic selectivity of nanopores under all three fields, especially in cases of hydrostatic pressure. By changing the electrolyte type, different ionic diffusion coefficients can significantly modulate the ionic selectivity under the concentration gradient. However, under electric fields and hydrostatic pressure, the ionic selectivity has little correlation with the ion species.

Based on our systematical investigation of ion transport and ionic selectivity of nanopores under different fields, we believe that the results can provide useful guidance for the design and application of high-performance devices for energy conversion based on nanoporous membranes.

**CRediT authorship contribution statement**

**Chao Zhang**: Data curations, software, visualization, and writing – original draft. **Mengnan Guo**: Data curations, validation, and writing – original draft. **Hongwen Zhang**: Writing – review & editing. **Xiuhua Ren**: Supervision. **Yinghao Gao**: Data curations. **Yinghua Qiu**: Investigation, methodology, conceptualization, funding acquisition, writing – original draft, and writing – review & editing.

**Declaration of competing interest**

The authors declare that they have no known competing financial interests or personal relationships that could have appeared to influence the work reported in this paper.

**Data availability**

Data will be made available on request.

**Acknowledgements**

This research was supported by the National Natural Science Foundation of China (52105579), the Guangdong Basic and Applied Basic Research Foundation (2023A1515012931), the Instrument Improvement Funds of Shandong University Public Technology Platform (ts20230107), and the Qilu Talented Young Scholar Program of Shandong University.

**Appendix A. Supplementary material**

Supplementary data to this article can be found online.

# Ionic Selectivity of Nanopores: Comparison among Cases under the Hydrostatic Pressure, Electric Field, and Concentration Gradient

Chao Zhang,[1] Mengnan Guo,[1] Hongwen Zhang,[2] Xiuhua Ren,[1] Yinghao Gao,[1] and Yinghua Qiu[2]*

1. School of Mechanical and Electronic Engineering, Shandong Jianzhu University, Jinan, 250101, China
2. Key Laboratory of High Efficiency and Clean Mechanical Manufacture of Ministry of Education, National Demonstration Center for Experimental Mechanical Engineering Education, School of Mechanical Engineering, Shandong University, Jinan, 250061, China

*Corresponding author: yinghua.qiu@sdu.edu.cn

# 1. Additional simulation details

Table S1 Boundary conditions used in simulations. COMSOL Multiphysics was used to solve the coupled Poisson–Nernst–Planck and Navier–Stokes equations.

| Scheme | Surface | Poisson | Nernst-Planck | Navier-Stokes |
|---|---|---|---|---|
| 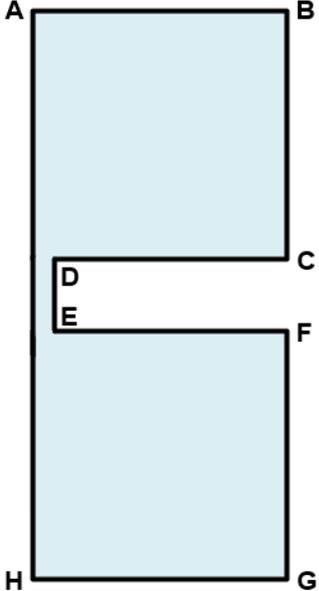 Voltage Gradient | AB | Constant potential $\phi = 0$ | Constant concentration $c_i = C_{bulk}$ | Constant pressure $p = 0$ No viscous stress $\mathbf{n}\cdot[\mu(\nabla\mathbf{u}+(\nabla\mathbf{u})^T)]=0$ |
| | BC, FG | Neutral surface $-\mathbf{n}\cdot(\varepsilon\nabla\phi)=0$ | No flux $\mathbf{n}\cdot\mathbf{N}_i = 0$ | No slip $\mathbf{u} = 0$ |
| | CD, DE, EF | $-\mathbf{n}\cdot(\varepsilon\nabla\phi)=\sigma_w$ | No flux $\mathbf{n}\cdot\mathbf{N}_i = 0$ | No slip $\mathbf{u} = 0$ |
| | HG | Constant potential $\phi = V_{app}$ | Constant concentration $c_i = C_{bulk}$ | Constant pressure $p = 0$ No viscous stress $\mathbf{n}\cdot[\mu(\nabla\mathbf{u}+(\nabla\mathbf{u})^T)]=0$ |
| | AH | Axial symmetry | Axial symmetry | Axial symmetry |
| 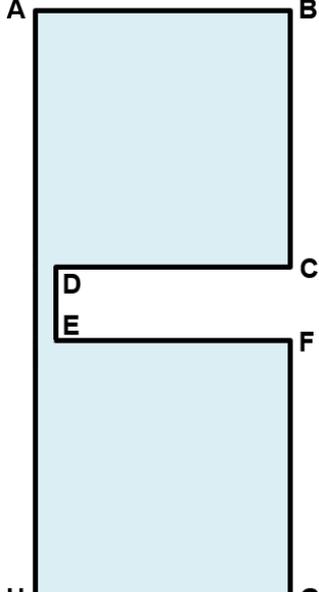 Pressure Gradient | AB | Constant potential $\phi = 0$ | Constant concentration $c_i = C_{bulk}$ | Constant pressure $p = 0$ No viscous stress $\mathbf{n}\cdot[\mu(\nabla\mathbf{u}+(\nabla\mathbf{u})^T)]=0$ |
| | BC, FG | Neutral surface $-\mathbf{n}\cdot(\varepsilon\nabla\phi)=0$ | No flux $\mathbf{n}\cdot\mathbf{N}_i = 0$ | No slip $\mathbf{u} = 0$ |
| | CD, DE, EF | $-\mathbf{n}\cdot(\varepsilon\nabla\phi)=\sigma_w$ | No flux $\mathbf{n}\cdot\mathbf{N}_i = 0$ | No slip $\mathbf{u} = 0$ |
| | HG | Constant potential $\phi = 0$ | Constant concentration $c_i = C_{bulk}$ | Constant pressure $p = P_{app}$ No viscous stress $\mathbf{n}\cdot[\mu(\nabla\mathbf{u}+(\nabla\mathbf{u})^T)]=0$ |
| | AH | Axial symmetry | Axial symmetry | Axial symmetry |
| AB | Constant potential $\phi = 0$ | Constant concentration $c_i = C_L$ | Constant pressure $p = 0$ No viscous stress $\mathbf{n}\cdot[\mu(\nabla\mathbf{u}+(\nabla\mathbf{u})^T)]=0$ | |
| BC, FG | Neutral surface $-\mathbf{n}\cdot(\varepsilon\nabla\phi)=0$ | No flux $\mathbf{n}\cdot\mathbf{N}_i = 0$ | No slip $\mathbf{u}=0$ | |
| CD, DE, EF | $-\mathbf{n}\cdot(\varepsilon\nabla\phi)=\sigma_w$ | No flux $\mathbf{n}\cdot\mathbf{N}_i = 0$ | No slip $\mathbf{u}=0$ | |
| HG | Constant potential $\phi = 0$ | Constant concentration $c_i = C_H$ | Constant pressure $p = 0$ No viscous stress $\mathbf{n}\cdot[\mu(\nabla\mathbf{u}+(\nabla\mathbf{u})^T)]=0$ | |
| AH | Axial symmetry | Axial symmetry | Axial symmetry | |

(Left panel shows Concentration Gradient schematic with vertices A, B, C, D, E, F, G, H.)

$\phi$, $V_{app}$, $\varepsilon$, $C_{bulk}$, $P_{app}$, $C_L$, $C_H$, $p$, $\mathbf{n}$, $\mathbf{N}_i$, $\mathbf{u}$, $\sigma_w$, $\mu$ are the surface potential, applied voltage, dielectric constant, bulk concentration, applied pressure, low bulk concentration, High bulk concentration, pressure, normal vector, flux of ions, fluid velocity, surface charge density of the pore wall and solution viscosity, respectively.

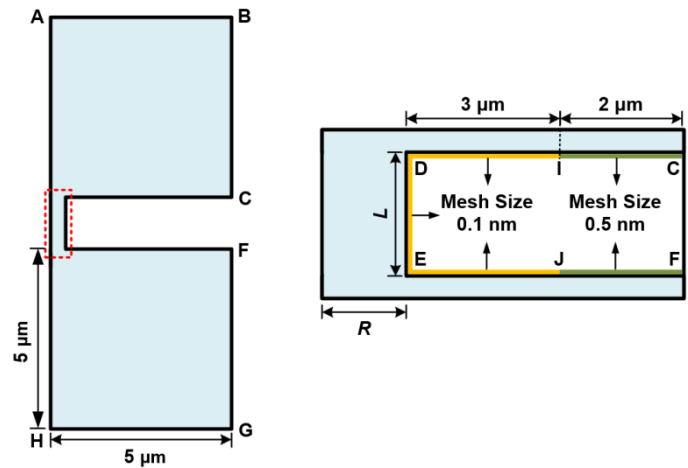

Fig. S1. Scheme of built mesh in simulations. For the inner-pore surface (DE) and 3 μm-wide regions of exterior surfaces beyond the pore boundary (DI and EJ), the mesh size of 0.1 nm was used. For the rest part of the exterior surfaces (IC and JF), the mesh size of 0.5 nm was chosen.

## 2. Additional simulation results

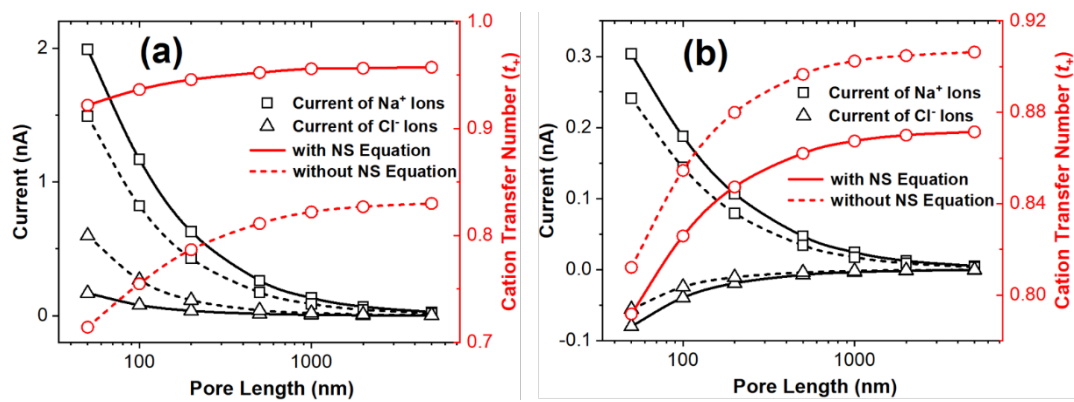

Fig. S2. The effects of electroosmotic flow (a) and diffusio-osmotic flow (b) on the ionic current and selectivity in nanopores of different pore lengths under the electric field and the concentration gradient, respectively. The pore diameter was 10 nm and the pore length varied from 50 to 5000 nm. A voltage of 0.6 V was used under the electric field. A concentration gradient of 0.01: 0.5 M NaCl was applied across nanopores under the concentration gradient. The surface charge density of pore walls was −0.08 C/m$^2$.

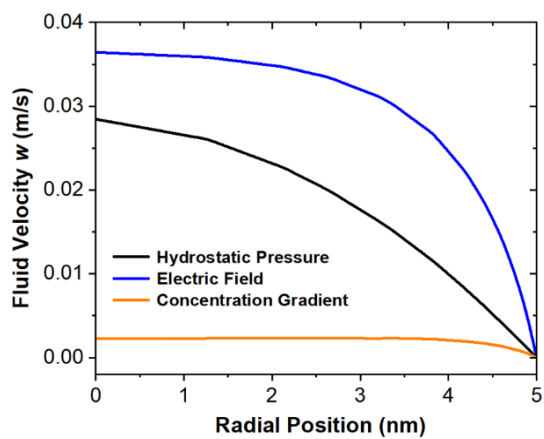

Fig. S3. Radial distributions of fluid velocity in the central cross-section of the nanopore under the hydrostatic pressure, electric field, and concentration gradient. The pore length and diameter were 1000 and 10 nm, respectively. A hydrostatic pressure of 4 MPa and a voltage of 0.6 V were applied across the nanopore under the hydrostatic pressure and electric field, respectively. 0.1 M NaCl solution was used inside reservoirs and nanopores in the above cases. A concentration gradient of 0.01: 0.5 M NaCl solution was applied across the nanopore under the concentration gradient. The surface charge density of pore walls was −0.08 C/m$^2$.

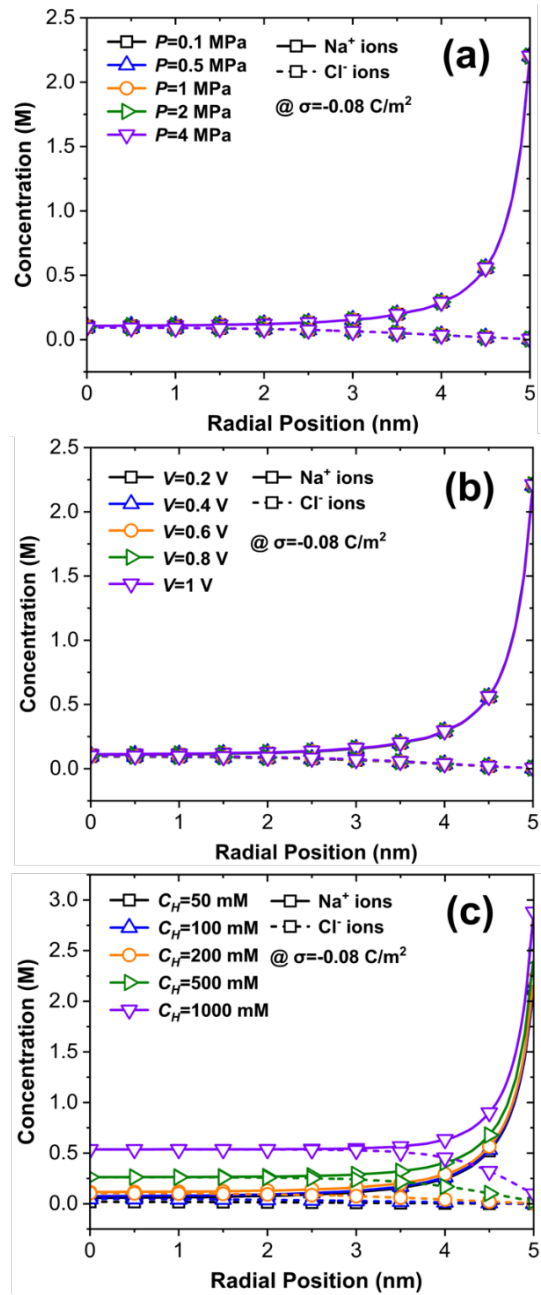

Fig. S4. Radial distributions of ion concentrations in the central cross-section of the nanopore under hydrostatic pressure (a), electric fields (b), and concentration gradients (c). The diameter and length of nanopores were 10 nm and 1000 nm. The legend of Fig. (c) denoted the concentration of the high-concentration side ($C_H$).

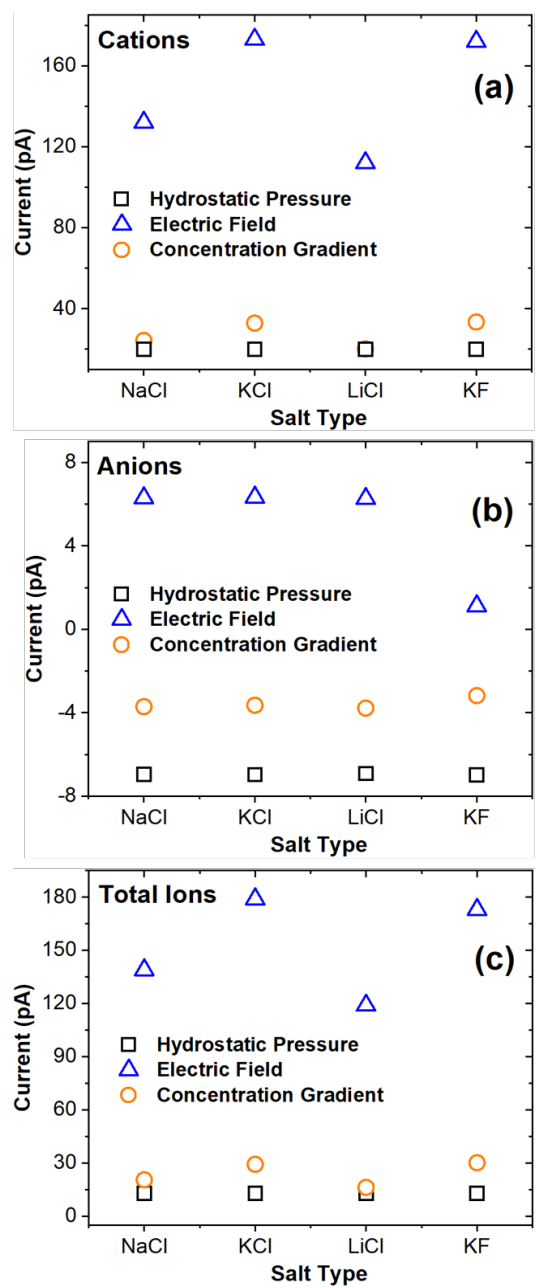

Fig. S5. Ionic current contributed by cations (a), anions (b), and total ions (c) in different electrolyte solutions under three fields, i.e. the hydrostatic pressure, electric field, and concentration gradient. The pore length and diameter were 1000 and 10 nm, respectively. A hydrostatic pressure of 4 MPa and a voltage of 0.6 V were applied across the nanopore under the hydrostatic pressure and electric field, respectively. 0.1 M electrolyte solutions were used inside reservoirs and nanopores in the above cases. A concentration gradient of 0.01: 0.5 M electrolyte solutions were applied across the nanopore under the concentration gradient. The surface charge density of pore walls was −0.08 C/m$^2$.

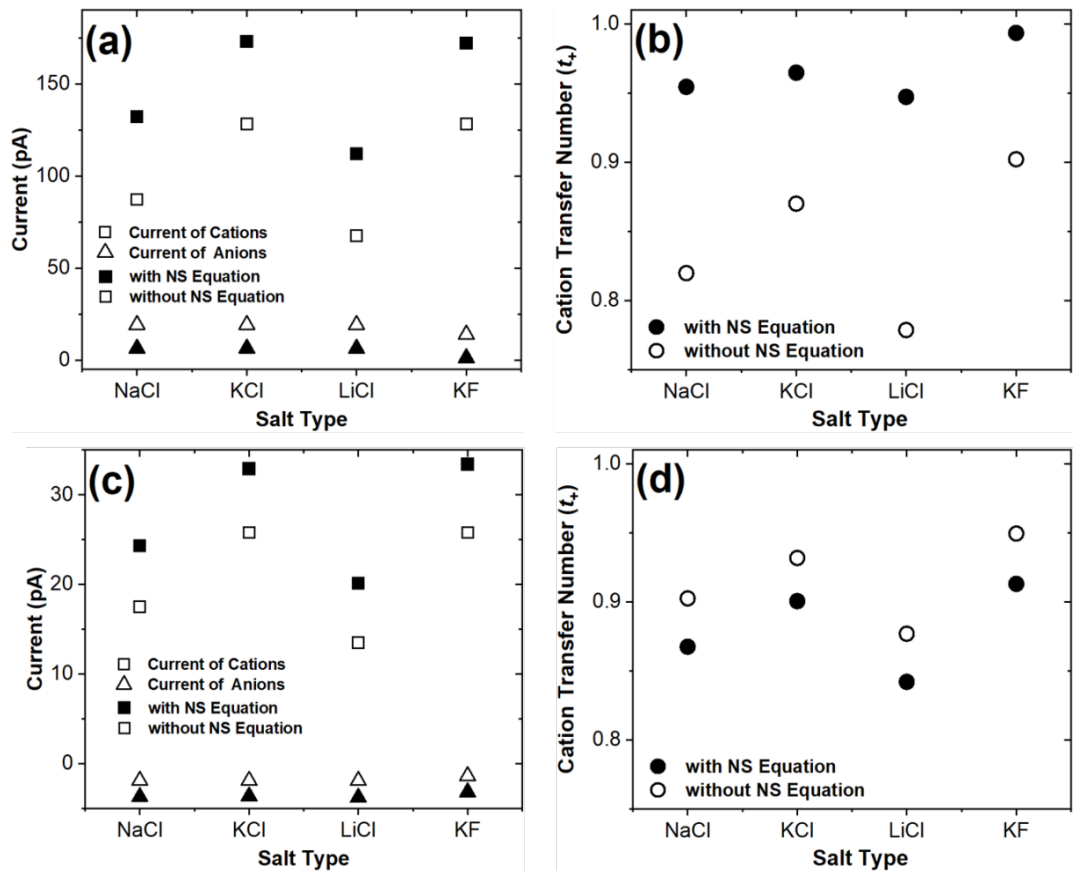

Fig. S6. The effects of electroosmotic flow (a-b) and diffusio-osmotic flow (c-d) on the ionic current and selectivity in nanopores under the electric field and the concentration gradient, respectively. In the simulations without NS equations, the fluid flow was not considered. Different electrolyte solutions were considered. 1000 and 10 nm were used in the pore length and diameter. A voltage of 0.6 V and a concentration gradient of 0.01: 0.5 M electrolyte solutions were applied across nanopores under the electric field and concentration gradient, respectively. 0.1 M electrolyte solutions were used inside reservoirs and nanopores under the electric field. The surface charge density of pore walls was −0.08 C/m$^2$.